\documentclass[pra]{revtex4}
\usepackage{times,amsfonts}
\usepackage{graphicx}
\usepackage{amsmath}
\usepackage{amssymb}
\usepackage{hyperref}
\usepackage{bm}
\usepackage{bbm}
\begin{document}
\title{Ultracold Gases of Ytterbium:  Ferromagnetism and  Mott States in an SU(6) Fermi System} 
\author{M.~A. Cazalilla}
\affiliation{Centro de F\'{\i}sica de Materiales (CFM). Centro Mixto CSIC-UPV/EHU. 
Edificio Korta, Avenida de Tolosa, 72. 20018 San Sebasti\'an. Spain}
\affiliation{Donostia International Physics Center (DIPC), Manuel de Lardiz\'abal 4, 20018 San Sebasti\'an, Spain.}
\author{A.~F.~ Ho}
\affiliation{Department of Physics, Royal Holloway, University of London, Egham, Surrey
TW20 0EX, UK.}
\author{M. Ueda}
\affiliation{ERATO Macroscopic Quantum Control Project, JST, Yayoi, Bunkyo-Ku, Tokyo 113-8656, Japan.}
\affiliation{Department of Physics, University of Tokyo, Hongo, Bunkyo-ku, Tokyo 113-0033, Japan.}
\begin{abstract} 
It is argued that ultracold quantum degenerate gas of ytterbium $^{173}$Yb atoms 
having nuclear spin $I = 5/2$ exhibits an enlarged  SU$(6)$ symmetry. Within the Landau 
Fermi liquid theory, stability criteria against Fermi liquid 
(Pomeranchuk) instabilities in the spin channel are considered. Focusing on the SU$(n > 2)$ generalizations
of ferromagnetism, it is shown within mean-field theory that the transition from the paramagnet to the itinerant 
ferromagnet is generically first order.  On symmetry grounds, general SU$(n)$ itinerant ferromagnetic ground states  
and their topological excitations are also discussed.  These SU$(n > 2)$ ferromagnets can become stable by increasing the scattering length using optical methods or in an optical lattice. However, in an optical lattice at current experimental
temperatures, Mott states with different filling are expected to coexist in the same trap, 
as obtained from a calculation based on the SU$(6)$ Hubbard model.
\end{abstract}

\date{\today}
\maketitle

\section{Introduction}

 Recently, the Kyoto group has managed to cool down to quantum degeneracy
 five Ytterbium isotopes~\cite{Kyoto_exp}.  The Ytterbium atom 
 has a closed-shell electronic structure in the ground state  ([Xe] $4$f$^{14}$ $5$s$^2$ 
 $^1$S$_0$), and hence its spin stems \emph{entirely} from the nuclear
 spin, $I$. The case of the fermionic species $^{173}$Yb is particularly interesting, 
 as it has nuclear spin $I = F = 5/2$. Hence $2F+1 = n = 6$, and the atom can be in
six different internal states. At ultracold temperatures,  experiments show that
the scattering length is independent of the atom internal state~\cite{Takahashi}.
This can be understood from the absence of electronic spin in the atomic ground state, 
and the extremely weak dependence of the inter-atomic potential on the atomic nuclear spin.
Thus, whereas for a spin-$5/2$ fermion the Lee-Yang-Huang  pseudo-potential,
depends  on three scattering lengths~\cite{YipHo99},  $a_s^{F= 0, 2, 4}$, the 
previous observation implies that $a_s^{0} = a_{s}^{2} = a_{s}^{4} = a_s$. Mathematically,
the interaction part of the Hamiltonian becomes:
\begin{equation}
H_{\mathrm{int}} = \frac{4\pi \hbar^2}{M} \sum_{i < j=1}^{N} \left[ a^0_s \mathcal{P}_0(ij) + a^2_s \mathcal{P}_2(ij)  + a^4_s
\mathcal{P}_4(ij)\right]\: \delta(\mathbf{r}_i-\mathbf{r}_j) = \frac{4\pi \hbar^2 a_s}{M} \sum_{i < j=1}^{N}\delta(\mathbf{r}_i-\mathbf{r}_j),
\end{equation}
where $M$ is the atom mass and $\mathcal{P}_F(ij)$ the projector onto the state of total
spin $F$ for the pair of particles $i$ and $j$.  Therefore,  the kinetic and interaction terms have the same
symmetry, that is, the initial SU$(2)$ spin-symmetry of the Hamiltonian 
describing an ultracold gas of  $^{173}$Yb atoms is enlarged to an 
effective SU$(6)$ symmetry.This is particularly interesting since  enlarged symmetries 
usually lead to additional spectral degeneracies~\cite{su6_baryon}, 
which  in turn can lead to exotic (correlated) ground states  and topological 
excitations~\cite{so5_papers,honerkamp_sun,Guan08,qhferromagnet}. 

The occurrence of SU$(6)$ in 
an ultracold gas  can also lead to new and interesting 
connections with high-energy physics, where SU$(6)$ has
been used to describe the flavor symmetry of  spinful quarks, 
as nuclear forces seem to be
spin independent to a first approximation~\cite{su6_baryon}. 
Indeed, some of the phases discussed below 
can be regarded as (non-relativistic) pion condensates  that spontaneously break  SU$(6)$. 
In addition, these phases also bear some resemblance to the  
quantum Hall ferromagnets~\cite{qhferromagnet} discussed in two-dimensional electron gases 
with valley symmetry (such as graphene).  Their possible existence in ultracold gases of  $^{173}$Yb
can allow for larger control thanks to the large tunability of these atomic systems. In this regard, ultracold 
$^{173}$Yb  atoms in optical lattices may also allow  the observation of  other exotic time-reversal 
symmetry breaking phases such as the staggered flux phase~\cite{Marston}, 
which has been speculated as the explanation to the anomalous properties of the pseudo-gap phase of 
the  high-$T_c$ cuprate  superconductors~\cite{SF_papers}.

 In this paper, we study (in Sect.~\ref{sec:FLT}) the Fermi liquid instabilities in the spin SU$(n = 6)$ 
channel of a strongly interacting $^{173}$Yb gas.  Focusing mainly on ferromagnetism,
which breaks the SU$(n)$ symmetry but not the space rotation invariance, we find in Sect.~\ref{sec:FM}  
that the paramagnetic to ferromagnetic transition to be generically first order for $n > 2$ at the mean-field level.  
On physical and symmetry grounds, we also identify the possible  broken-symmetry ground states. The possibility of 
spontaneously breaking the SU($n$) symmetry group in  a cascade of phase transitions between different ferromagnetic 
phases  hints at  a much richer   phase diagram than in the spin-$\frac{1}{2}$ case~\cite{FM_papers}.   
These phases will also sustain exotic topological excitations, such as skyrmions in two dimensions
and monopoles in three dimensions. As argued below, these phenomena may
be observed by increasing the scattering length using an optical Feshbach 
resonance~\cite{optical_feshbach} or in perhaps also in a deep optical lattice. In Sect.~\ref{sec:HM},
we consider the situation in the lattice. Close to half-filling, \emph{i.e.} 3 atoms per site, 
many phases, which may~\cite{honerkamp_sun} or may not~\cite{Marston}
break the SU$(6)$ group, are likely to exist. However, at current accessible optical-lattice temperatures, 
atom hopping is largely incoherent, and  Mott states are likely to coexist in the same harmonic trap. 
Indeed, for an SU$(6)$ Hubbard model at high temperatures, we have computed the density profile
showing the Mott plateaux (see~Fig.~\ref{fig:fig1}). Finally, a summary of the results as well as 
a brief discussion of how to detect some of the phases discussed here can be found in
Sect.~\ref{sec:concl}.

\section{SU$(n)$ Fermi liquid and Fermi surface instabilities}\label{sec:FLT}

 We begin our analysis of the $^{173}$Yb system,  
 by exploring some consequences of SU$(n=6)$ for the Fermi liquid phase of 
 an interacting gas of $^{173}$Yb atoms. Although we shall focus on  the continuum case,
 many of there results in this section can be readily applied to the Fermi liquid phase
 of  the gas loaded in an optical lattice   (we neglect harmonic confinement for the moment;
 it will be considered briefly at the end, and more thoroughly elsewhere~\cite{unpub}). 
Following Landau~\cite{BaymPethick}, 
we describe the low-lying excited states of the system using the distribution function 
$n^{\alpha}_{\beta}(\mathbf{p}) = \langle \psi^{\dag}_{\beta}
(\mathbf{p}) \psi^{\alpha}(\mathbf{p})\rangle$ of a set of elementary 
excitations called Landau quasi-particles (QP, essentially atoms `dressed' by the 
interactions). The latter are annihilated (created) by the Fermi operator 
$\psi^{\alpha}(\mathbf{p})$ ($\psi^{\dag}_{\alpha}(\mathbf{p})$) 
carrying (lattice) momentum $\hbar \mathbf{p}$ and  
SU$(n)$ index $\alpha = 1, \ldots, n$. 
The excitation free energy of the QP states  is given by the  
Landau functional (summation over repeated Greek-indices is implied henceforth):
\begin{align}
\delta F &= \sum_{\mathbf{p}} \left[\varepsilon^{0}(\mathbf{p}) -\mu\right] 
\delta n^{\alpha}_{\alpha}(\mathbf{p}) 
+ \frac{1}{2\Omega}\sum_{{\bf p},{\bf p}'} f^{\alpha\beta}_{\gamma \delta}(\mathbf{p},\mathbf{p}') 
\delta n^{\gamma}_{\alpha}(\mathbf{p}) \delta n^{\delta}_{\beta}(\mathbf{p}'),
\end{align}
where $\Omega$ is the system volume, $\epsilon^{0}(\mathbf{p})$ is the excitation energy of a single 
Landau quasi-particle carrying momentum $\hbar \mathbf{p}$. We assume 
the ground state to be an SU$(n)$ singlet and therefore the ground
state quasi-particle distribution $[n^{0}]^{\alpha}_{\beta}(\mathbf{p})= 
\theta(\mu -\epsilon^{0}(\mathbf{p}))\: \delta^{\alpha}_{\beta}$, where $\mu$ is the chemical
potential; $\delta n^{\alpha}_{\beta}(\mathbf{p}) = 
n^{\alpha}_{\beta}(\mathbf{p})-[n^{0}]^{\alpha}_{\beta}(\mathbf{p})$. The Landau
functions $f^{\alpha\beta}_{\gamma\delta}(\mathbf{p},\mathbf{p}') = 
f^{\beta\alpha}_{\delta\gamma}(\mathbf{p},\mathbf{p}') $ describe  
interactions between quasi-particles.  The expression for
$\delta F$ can be considerably simplified with the help of group
theory by noticing that $\delta n^{\alpha}_{\beta}(\mathbf{p})$ transforms
as a tensor belonging to the \emph{reducible} representation of SU$(n)$
$n\otimes \bar{n} = 1 \oplus (n^2-1)$, where $n$ and $\bar{n}$ are the
fundamental and its complex conjugate representations, whereas  
$1$ is the singlet and $n^2-1$ the adjoint representations, respectively.
Therefore, $\delta n^{\alpha}_{\beta}(\mathbf{p}) = 
\frac{1}{n}\delta \rho(\mathbf{p})\: \delta^{\alpha}_{\beta}  
+ \sum_{a=1}^{n^2-1}\delta m^{a}(\mathbf{p}) \left(\mathbbm{T}^{a}\right)^{\alpha}_{\beta}$,
where $\mathbbm{T}^a$ are the (traceless) generators of the SU$(n)$ Lie-algebra
obeying $\left[ \mathbbm{T}^{a},\mathbbm{T}^{b} \right] = i \sum_{c=1}^{n^2-1}\lambda^{abc} \: \mathbbm{T}^c$;
choosing the normalization such that ${\rm Tr}  \left( \mathbbm{T}^{a}\mathbbm{T}^{b}\right) = \frac{1}{2}
\delta^{ab}$, the structure constants $\lambda^{abc}$ are fully anti-symmetric.
In this representation, $\delta\rho(\mathbf{p})$ describes the total density fluctuations 
and $\delta m^a(\mathbf{p})$ the SU$(n)$ magnetization fluctuations.
In addition, since the Landau functions transform as tensors belonging to 
$n\otimes n\otimes \bar{n} \otimes \bar{n} = 1 \oplus 1 \oplus$  non-singlet representations,
all the tensor components are determined by just \emph{two} scalar functions (compared to
the five needed for a $F = 5/2$ Fermi gas~\cite{YipHo99}), that is, 
$f^{\alpha\beta}_{\gamma\delta}(\mathbf{p},\mathbf{p}') = f^{\rho}(\mathbf{p},\mathbf{p}') \delta^{\alpha}_{\gamma}
\delta^{\beta}_{\delta} + 2 f^{m}(\mathbf{p},\mathbf{p}')\sum_{a=1}^{n^2-1} \left(\mathbbm{T}^a \right)^{\alpha}_{\gamma} 
\left(\mathbbm{T}^{a}\right)^{\beta}_{\delta}$. 

  We next consider  the stability of the Fermi surface (FS) of  the SU$(n)$ Fermi liquid just described above.
For an isotropic FS,  general stability conditions against FS deformations and pairing  were obtained  using  the renormalization group by Chitov and Senechal~\cite{CitovSenechal}. They concluded 
that pairing occurs for attractive interactions. For repulsive interactions, $d$-wave pairing is also 
possible on a lattice near half-filling, but the pairing temperature rapidly decreases
with increasing $n$~\cite{honerkamp_sun}.  In the case of $^{173}$Yb,  
s-wave interaction between  atoms is naturally repulsive, as
the scattering length is $a_s = +10.55$ nm~\cite{Takahashi}.  This 
yields $p_F a_s \simeq 0.1$ at the center of the trap in 
current experimental conditions~\cite{Kyoto_exp,Takahashi_private}.
Furthermore,  currently accessible temperatures $T/\mu \simeq 0.4$~\cite{Kyoto_exp,Takahashi_private} 
are well above  any pairing temperature scale. Therefore, the Fermi liquid phase should be a good starting 
description of the system.  However, if the interaction is made sufficiently repulsive, the Fermi liquid   can 
become unstable.  The stability of the FS to the so-called Pomeranchuk instabilities
can be assessed within Fermi liquid theory   
by considering the excitation energy of a quasi-particle distribution describing a deformation 
of the FS~\cite{BaymPethick}. In
matrix notation:  $\mathbbm{n}(\mathbf{p})=\theta \left[ \left(\mu - \epsilon^{0}(\mathbf{p})\right) \mathbbm{1}+ 
\delta\mathbbm{u}(\mathbf{\hat{p}})\right]$, where $\mathbbm{n}(\mathbf{p})$ denotes the matrix whose
components are $n^{\alpha}_{\beta}(\mathbf{p})$, $\mathbbm{1}$ is the unit matrix, and 
$\delta \mathbbm{u}(\mathbf{\hat{p}})$ is a matrix  function that describes a small local
deformation of the FS ($\mathbf{\hat{p}}$ denotes those $\mathbf{p}$-points lying on the FS). 
Expanding in powers of  $\delta\mathbbm{u}(\mathbf{\hat{p}}) = \frac{1}{n}\delta u_{\rho}(\mathbf{\hat{p}}) \mathbbm{1}
+ \sum_a \delta u^a_{m}(\mathbf{\hat{p}}) \mathbbm{T}^{a}$
up to second order, we obtain $\delta \rho(\mathbf{p})  = \delta \left(\mu-\epsilon^0(\mathbf{p}) \right) \delta u_{\rho}({\mathbf{\hat{p}}}) + \frac{1}{2!} \delta^{\prime}
\left(\mu-\epsilon^0(\mathbf{p})\right) \left[ \frac{1}{n} \left( \delta u_{\rho}(\mathbf{\hat{p}})\right)^2 +  
\frac{1}{2}\: \sum_{a} \left( \delta u^a_{m}(\mathbf{\hat{p}})\right)^2\right]+ \cdots$ and
$\delta m^a(\mathbf{p}) = \delta \left(\mu-\epsilon^0(\mathbf{p}) \right) \delta u^{a}_{m} (\mathbf{\hat{p}})+ \cdots$ 
In the continuum  or  in an optical lattice at  low-filling, the FS is isotropic
at the locus where $|\mathbf{p}| = p_\mathrm{F}$, where $p_\mathrm{F}$ is the Fermi momentum,
and  $\mathbf{p}/|\mathbf{p}| = \mathbf{\hat{p}}$.
For example, in three dimensions, we can expand
$\delta u_{\rho}(\mathbf{\hat{p}})  = \sum_{LM} \delta u^{LM}_{\rho}\: Y_{LM}(\mathbf{\hat{p}})$,
and  $\delta u^{a}_{\rho}(\mathbf{\hat{p}})  = \sum_{LM} \delta u^{a,LM}_{m}\:  Y_{LM}(\mathbf{\hat{p}})$,
$f^{\rho/m}(\mathbf{p},\mathbf{p}') =  f^{\rho/m}(\mathbf{\hat{p}}\cdot\mathbf{\hat{p}}') 
 = \frac{4\pi}{2L+1}\sum_{LM} f^{L}_{\rho/m} 
Y^{*}_{LM}(\mathbf{\hat{p}}) Y_{LM}(\mathbf{\hat{p}'})$, where $Y_{LM}(\mathbf{\hat{p}})$ are the 
spherical harmonics. Therefore, $\delta F = \delta F_{\rho} + \delta F_{m}$, where
(note that $\delta u^{LM}_{\rho}, \delta u^{a,LM}_{m}$ have units of energy)
\begin{align}
\delta F_{\rho} &= \frac{\Omega N^{0}(\mu)}{8\pi n} \sum_{LM} \left[ 1 +  \frac{\mathcal{F}_{L}^{\rho}}{2L+1}\right]\: \left|\delta u^{LM}_{\rho}\right|^2,\label{eq:srho}\\
\delta F_{m} &= \frac{\Omega N^{0}(\mu)}{16\pi} \sum_{LM} \left[   1 +  \frac{\mathcal{F}_{L}^{m}}{2L+1} \right]\: 
 \sum_{a} \left|\delta u^{a,LM}_{m}\right|^2.\label{eq:sm}
\end{align}
We have introduced the (dimensionless) Landau parameters defined as $\mathcal{F}_{L}^{\rho} = 
n N^0(\mu) f_{L}^{\rho}$ and $\mathcal{F}_L^{m} = N^{0}(\mu) f_{L}^m$, where
$N^{0}(\mu) = M^* p_F /(2\pi^2 \hbar^2)$ is the quasi-particle density of states (per species) at the FS in three dimensions, 
$M^*$ being the quasi-particle effective mass and $\hbar p_F $ the Fermi momentum ($p_F =  (6 \pi^2 \rho_0/n)^{1/3}$, where
$\rho_0$ is the total density).   Hence, from Eqs.~(\ref{eq:srho}) and (\ref{eq:sm}), the FS will be unstable if $\mathcal{F}^{L}_{\rho/m} <  -(2L+1)$, for $L=0,1,\ldots$

 The FS instabilities in the density channel ($\delta F_{\rho} < 0$, that is,
$\mathcal{F}^{\rho}_L < -(2L+1)$) are 
formally identical  to those occurring in Fermi systems with no spin.
They have received much attention recently~\cite{Pomeranchuk},
and lead to phases where  the rotation (or point-group, in the lattice)  symmetry of the FS is broken
($L>0$).    On the other hand,  much less attention has focussed on instabilities in the spin channel, 
which also break spin symmetry~\cite{Pomeranchuk_spin}, and may occur in the interesting
case of the $^{173}$Yb system with SU$(n=6)$ symmetry.
Certainly,  the most exotic states will be those resulting from an  instability with
 $L > 0$  in the spin channel, or, for a non-isotropic FS, one that  breaks the lattice point-group  besides 
 SU$(n)$. The resulting states have a much more complex 
order parameter, $\Phi^a_{LM} \propto  \int d\mathbf{\hat{p}} \: Y_{LM}(\mathbf{\hat{p}}) 
\, (\mathbbm{T}^{a})^{\beta}_{\alpha}  \delta n^{\alpha}_{\beta}(p_\mathrm{F} \mathbf{\hat{p}})$, 
that is the product of an orbital
and an SU$(n)$ part (similar to superfluidity in $^{3}$He). 

  However, as ultracold $^{173}$Yb atoms naturally interact via \emph{repulsive} 
s-wave (contact)  interactions, in the isotropic case  $\mathcal{F}_{0}^{m}$ is
expected to be the most negative Landau parameter, thus favoring SU$(6)$ ferromagnetic 
correlations. Indeed,   within Hartree-Fock theory, $\mathcal{F}_{0}^{\rho} =  N_{0}(\mu) g(n-1) > 0$ 
whereas $\mathcal{F}_{0}^{m} = - N_{0}(\mu) g < 0$ 
($N_{0}(\mu) = M p_F/(2\pi^2 \hbar^2)$, where $M$ is the atom mass,  $g = 4\pi \hbar^2 a_s/M$.   
Hence, the SU$(n)$  generalization of  Stoner's  criterion for ferromagnetism,  $F^{m}_{0} < -1$, yields $N_0(\mu) g > 1$, or
equivalently, $p_F a_s >  p_F a^*_s  = \frac{\pi}{2}$, in the continuum case.  It is worth noting that
this criterion turns out to be the same as for the SU$(2)$ case, that is, 
it is independent of $n$. The independence on $n$ of the Stoner criterium in a SU$(n)$ Fermi
system  can be understood using a simple energetic  argument: To
create a polarized ground state, imagine for example that  $\delta M_0/(n-1)$ fermions are removed from 
the Fermi surface of each of  the  $n-1$ flavors with  $\alpha <  n$ and added 
to the Fermi surface of the $\alpha = n$ flavor (so that the total particle number is unchanged). 
For small $\delta M_0$, the  kinetic energy of the system increases  by 
$ (n-1) \frac{\left[\delta M_0/(n-1)\right]^2}{2 \Omega N_0(\mu)} + \frac{\left( \delta M_0\right)^2}{2 \Omega N_0(\mu)} 
= \frac{n (\delta M_0)^2}{2 (n-1) \Omega N_0(\mu)}$, whereas the (Hartree-Fock) interaction energy decreases by
$\frac{g}{\Omega} \left[ \frac{(n-1)(n-2)}{2} \left(\frac{\delta M_0}{n-1}\right)^2 
- (n-1) \left( \frac{\delta M_0^2}{n-1} \right) \right] = - \frac{n g (\delta M_0)^2}{2 (n-1) \Omega}$. Hence, upon
comparing both energies, the dependence on $n$ drops out and the system becomes 
unstable provided that $N_0(\mu) g > 1$, which is independent of $n$ and agrees with
the result obtained from Fermi liquid theory in the Hartree-Fock approximation. The cancellation of the dependence 
on $n$ to the lowest order is a consequence of the fact that both the kinetic and exchage energies scale linearly with $n$
(inspite of the fact that, na\"ively, the interaction scales as $n^2$).  Nevertheless, as we shall see below, the nature of the transition
from a paramagnet to an itinerant ferromagnet  turns out to be very different for SU$(n)$
with $n > 2$.

\section{SU$(n)$ itinerant ferromagnets}\label{sec:FM}

The previous analysis using Landau Fermi liquid theory  does not tell us
anything about the order of the transition. In the spin-$\frac{1}{2}$ (SU$(2)$) case, a Landau 
free-energy functional  obtained from the microscopic Hamiltonian finds 
a continuous transition~\cite{Hertz76, FM_papers}. However,  it has been recently pointed out 
that the coupling of the order parameter fluctuations  to soft modes changes the order of 
the transition from second to first order at  low temperatures~\cite{KB05,FM_papers}. 
In order to gain further insights into the nature of the transition at the mean field level, we shall
derive  in this section an effective action for the ferromagnetic order parameter starting from the microscopic
model. To this end, we use the following operator identity for the interaction term of the Hamiltonian
density:
\begin{equation}
{\cal H}_{\mathrm{int}}(\mathbf{r})  = \frac{1}{2} g \, \bar{c}_{\alpha}(\mathbf{r}) \bar{c}_{\beta}(\mathbf{r})
c^{\beta}({\bf r}) c^{\alpha}({\bf r}) = \frac{(n-1)}{2n} g \, :\left[ \rho(\mathbf{r}) \right]^2: - g \sum_{r=2}^{n} :\left[ \bar{c}_{\alpha}(\mathbf{r})  \: ( \mathbbm{T}^{r^2-1})^{\alpha}_{\beta}\,  c^{\beta}(\mathbf{r}) \right]^2 : 
\end{equation}
In the above expression $:\ldots:$ stands for operator normal order, that is, the prescription that
all atom creation fields, $\bar{c}_{\alpha}(\mathbf{r})$, should stand to the left of the
destruction fields, $c^{\alpha}(\mathbf{r})$. The matrices $\mathbbm{T}^{r^2-1} = 
\frac{1}{\sqrt{2 r (r-1)}} \mathrm{diag}\left(1, 1, \ldots, 1-r,\ldots, 0,0 \right)$ are the diagonal
generators of the Lie algebra (\emph{i.e.} the Cartan subalgebra). We next perform
a Hubbard-Stratonovich decoupling of the density ($\propto \rho^2$) and SU$(n)$-spin 
interaction terms, which yields the following action ($\beta = 1/T$, $T$ being the absolute temperature)
\begin{align}
S[\bar{c}_{\alpha}, c^{\alpha}, \varphi, \{ \phi_r \}]  &= \int d\mathbf{r} \int^{\hbar \beta}_0 \frac{d\tau}{\hbar}\,
  \Big\{  \bar{c}_{\alpha}(\mathbf{r},\tau) \left[  \left( \hbar\partial_{\tau} - \mu - \frac{\hbar^2}{2M} \nabla^2 
  +  \frac{n-1}{n} \: g \varphi(\mathbf{r},\tau)   \right) \delta^{\alpha}_{\beta}  \right.
  \nonumber\\ 
  &\left.  - g \sum_{r=2}^{n} \mathcal{M}_r(\mathbf{r},\tau) (\mathbbm{T}^{r^2-1})^{\alpha}_{\beta}  \right] c^{\beta}(\mathbf{r},\tau)   
   - \frac{n-1}{2n} g \varphi^2(\mathbf{r},\tau) +  \frac{g}{4}  \sum_{r=2}^n  \mathcal{M}^2_r(\mathbf{r},\tau)  \Big\}. 
\end{align}
Following the work by Hertz~\cite{Hertz76} for the SU$(2)$ case, we focus on the 
SU$(n)$ spin fluctuations and therefore obtain an effective action for the fields $\mathcal{M}_r(\mathbf{r},\tau)$
by  integrating out the Fermions and setting the density-fluctuation field $\varphi(\mathbf{r},\tau)  
= \rho_0$ ($\rho_0$ being the total density), that is,  its saddle point value.  Such a procedure  yields the following effective action:
\begin{equation}
S_{\mathrm{eff}}[\mathbbm{M}] =  
- \mathrm{Tr} \ln \left[ - G^{-1}_0 \mathbbm{1} -  g \mathbbm{M} \right] + \frac{g}{2\hbar} \int d\mathbf{r} d\tau  \,  \mathrm{Tr} \, \mathbbm{M}^2(\mathbf{r},\tau)  \label{eq:seff}
\end{equation}
where $\mathbbm{M}(\mathbf{r},\tau) = \sum_{r=2}^{n} \mathcal{M}_r (\mathbf{r},\tau) \mathbbm{T}^{r^2-1}$, such that
$\mathrm{Tr}\, \mathbbm{M}(\mathbf{r},\tau) = 0$, and $G^{-1}_0(\mathbf{r}-\mathbf{r}',\tau-\tau') = -
\left(\hbar \partial_{\tau} - \mu + \frac{(n-1)}{n} g \rho_0 - \frac{\hbar^2}{2M} \nabla^2 \right)\delta(\mathbf{r}-\mathbf{r}) \delta(\tau-\tau')$.
However, it should be noticed that the present Hubbard-Stratonovich decoupling scheme using only the
\emph{ diagonal} generators of SU$(n)$ breaks the  full
SU$(n)$ invariance of the theory. Yet,  it does reproduce the correct Stoner criterion in the 
mean field (Hartree-Fock) approximation, which, as discussed above, comes out to be independent of $n$.  
The SU$(n)$ invariance  can be recovered by extending the functional integral
over the entire set of traceless hermitian  matrices $\mathbbm{M}$ transforming according to the
adjoint representation of SU$(n)$. As noted in Sect.~\ref{sec:FLT}, a
convenient of basis for this set is provided by the generators of the SU$(n)$ Lie algebra,
$\mathbbm{T}^a$, with $a = 1, \ldots, n^2-1$. Hence, $\mathbbm{M}(\mathbf{r},\tau) = \sum_{a} 
m_{a}(\mathbf{r},\tau)\, \mathbbm{T}^a$, where $m_a(\mathbf{r},\tau)$ are real fields. 
Near the paramagnetic-ferromagnetic phase transition,
we expect the order parameter to be small and therefore, we perform a series expansion
in $\mathbbm{M}$ neglecting its dependence in $\mathbf{r}$ and $\tau$. This yields the
following (Landau) free-energy per unit volume:
\begin{equation}
\frac{F}{\Omega} = \frac{F_0}{\Omega} + \sum_{n=2} \frac{g^n v_n}{n}  \mathrm{Tr} \, \mathbbm{M}^n = \frac{F_0}{\Omega} +  \frac{g^2}{2} v_2 \mathrm{Tr} \, \mathbb{M}^2
 + \frac{g^3}{3}  v_3 \mathrm{Tr} \, \mathbb{M}^3 + \frac{g^4}{4} v_4  \mathrm{Tr} \, \mathbb{M}^4 + \cdots \label{eq:lf}
\end{equation}
The term of $O(g)$ vanishes identically because $\mathbb{M}$ is traceless. However, for $n > 2$, 
terms of both even and odd order in $g$ are non zero and occur in the free energy expansion in powers of $\mathbbm{M}$.
This is to be contrasted with the  SU$(n=2)$ case, where only terms of even order occur~\cite{Hertz76,KB05,FM_papers}. 
The coefficients $v_2 = (g^{-1} + \chi_2)$
and $v_n = \chi_n$ for $n > 2$, where $\chi_n =  \frac{(-1)^n}{\beta \Omega}\sum_{k} \left[ G_0(k) \right]^{n} = - \frac{1}{(n-1)!}\frac{\partial^{n-2}  N_0(\mu)}{\partial \mu^{n-2}}$, $k = (i\epsilon_n,\mathbf{k})$ and $G_0(k) = \left(i\epsilon_n -\epsilon(\mathbf{p}) +\mu \right)^{-1}$, where $\epsilon_n = \frac{2\pi}{\beta} (n+\frac{1}{2})$, $\epsilon(\mathbf{p}) = \frac{\hbar^2 \mathbf{p}^2}{2M}$, and we have shifted the chemical potential $\mu \to \mu - g (n-1)/n \rho_0$ to account for its renormalization due to interactions.

We could have obtained the above free energy based on symmetry considerations of the order parameter.
However, the microscopic approach allows us to relate the coefficients of the expansion to the model parameters. 
We next  set  $\mathbbm{M} = \sum_{a=1}^{n^2-1} m_a \mathbbm{T}^a$ in (\ref{eq:lf}) and use
the following SU$(n)$ identity  (see \emph{e.g.}~\cite{Greiner}),
\begin{equation}
\mathbbm{T}^{a} \mathbbm{T}^b = \frac{1}{2n} \delta^{ab} \mathbbm{1} +  \frac{1}{2}  \sum_{c=1}^{n^2-1}\left (d^{abc}  + i f^{abc} \right)  \mathbbm{T}^c,
\end{equation}
where the group structure constants $d^{abc}$ are fully symmetric and $f^{abc}$ fully anti-symmetric~\cite{Greiner}. For $n = 2$, $d^{abc} = 0$ but for $n > 2$ these structure constants are non-zero,  which has  important implications for the order of the paramagnetic-ferromagnetic  phase transition. In terms of the $m_{a}$ components of the order parameter, the free-energy reads:
\begin{equation}
\frac{F}{\Omega} = \frac{F_0}{\Omega} + \frac{g^2}{4} v_2 \sum_{a}\left(m^a \right)^2  + \frac{g^3}{12} v_3 \sum_{abc} d^{abc} m_a m_b m_c  
+  \frac{g^4 }{16} v_4\left\{ \frac{1}{n}\left[ \sum_{a} \left( m_a \right)^2\right]^2 + \frac{1}{2} \sum_{abcde} d^{abe}d^{cde} m_{a} m_{b}m_{c}m_{d} \right\} + 
\cdots
\end{equation}
The above expression shows explicitly that the Landau free energy contains a cubic term in the order parameter $m_{a}$, which
implies that, at  least at the mean field level, the transition from the paramagnetic to the ferromagnetic phase
is first order.  Thus, the system will exhibit hysteresis, and phase coexistence, with finite surface tension between the
ferromagnetic and paramagnetic phase. Furthermore, the entropy will undergo a finite jump  across the phase transition from the
paramagnet to the SU$(n)$ ferromagnet.   Moreover,  the gas parameter resulting from Stoner's criterion $p_F a^{*}_s = \frac{\pi}{2}$, which is the point where the quadratic coefficient vanishes,   is actually  \emph{larger} than the critical value of the gas parameter,
that is, $p_F a^c_s < p_F a^{*}_s = \frac{\pi}{2}$. The latter corresponds to the point where both the paramagnetic minimum  ($\mathbbm{M} = 0$)  and ferromagnetic minimum ($\mathbbm{M} \neq 0$)  have the same free energy. 
 
  To illustrate the general ideas presented above, we shall next consider the case of the smaller group SU$(3)$, which already 
contains essential ingredients of SU$(n > 2)$ ferromagnetism. The more complicated case of 
SU$(6)$ relevant to an unpolarized  mixture of $^{173}$Yb atoms will be studied elsewhere~\cite{unpub}.
However, it is worth saying that the SU$(3)$ case would correspond to an experiment where the system is  prepared as
a mixture containing an equal population of \emph{only three} of the six internal states~\footnote{Similarly, smaller
unitary groups with $n< 6$ would correspond to mixtures with smaller number of components.}~\footnote{Another
possibility is based on the observation that the Lie algebra of SU$(6)$ can be written as a direct product
of SU$(3)_{\mathrm{flavor}}\otimes$SU$(2)_{\mathrm{spin}}$, where `flavor' corresponds to the magnitude of 
the nuclear spin (that is, $\frac{1}{2},\frac{3}{2}, \frac{5}{2}$) and the `spin' index to the sign ($\pm \frac{1}{2}$, etc.).
Thus, we can choose to write the order parameter $\mathbbm{M} = \sum^{\prime}_{fs} m_{fs} 
\mathbbm{u}^f\otimes \mathbbm{v}^{s}$, where $\{\mathbbm{u}^{f}\} = \left\{ \mathbbm{1}_f, \frac{1}{2}\lambda^{1}, 
\ldots,\frac{1}{2}\lambda^{8} \right\}$ and $\{\mathbbm{v}^{s}\} =  \left\{ \mathbbm{1}_s, \frac{1}{2}\sigma^1, \frac{1}{2}\sigma^2, \frac{1}{2}\sigma^3  \right\}$, are the generators of the $SU(3)_{\mathrm{flavor}}$ and SU$(2)_{\mathrm{spin}}$ Lie algebras, and 
the prime in the summation means that the operator $\mathbbm{1}_{f} \otimes \mathbbm{1}_{s}$ should
be excluded. Assuming that the  SU$(3)_{\mathrm{flavor}}$ can be broken but  the SU$(2)_{\mathrm{spin}}$ cannot,
then $\mathbbm{M} = \frac{1}{2} \sum_{f} m_{f} \lambda^{f}\otimes \mathbbm{1}_s = \mathbbm{M}_f \otimes \mathbbm{1}_s$.
However, it should be notice that this way of breaking the symmetry may not be energetically favorable.}
Considering a three dimensional gas in the continuum, setting $\mathbbm{M} = 
 U^{\dag}\left( m_3 \mathbbm{T}^{3} + m_8 \mathbbm{T}^8 \right) U$, where $U \in \mathrm{SU}(3)$, 
and  using cyclic property of the trace along with 
the parametrization $m_{3} = \frac{(p_F a_s)^{-1}}{p^3_F}  \: \bar{m}_0 \: \cos \theta$ and $m_{8} =  \frac{(p_F a_s)^{-1}}{p^3_F} \bar{m}_0 \: \sin \theta$,    we arrive at the following expression  for the (dimensionless) free-energy at $T \ll \mu$:
\begin{align}
\frac{(F-F_0)/\mu}{p^3_F  \Omega} =  \frac{c_2}{2} \left(\frac{1}{p_F a_s} - \frac{2}{\pi}\right)\:   \bar{m}_0^2 - 
\frac{c_3}{3}   \bar{m}_0^3  \sin 3\theta + \frac{c_4}{4}  \bar{m}_0^4  \nonumber  
- \frac{c_5}{5} \bar{m}_0^5 \sin3\theta + \frac{c_6}{6} \bar{m}_0^6 (10-\cos 6 \theta) + \ldots
\end{align}
In the above expression, the numerical coefficients are $c_2 = 4\pi, c_3 = 8\pi/\sqrt{3}, c_4 = 16 \pi^2/3, c_5 = 320 \pi^3/27\sqrt{3}$, and  $c_6 = 16 \pi^4/9$.  Using the above expression up to sixth order, we can also obtain the shift in the critical gas parameter (relative to the Stoner value, $p_F a^*_s = \frac{\pi}{2}$): $(p_F a^c_s) - (p_F a^*_s)^{-1} \simeq 0.066$ or $1-a^c_s/a^*_s \simeq 0.094 \simeq 10\%$ at the the mean field level. Fluctuations are likely to decrease the critical value of the gas parameter even further from the Stoner value, and  may also change the character of the transition (see \emph{e.g.}~\cite{deGennes}). In the SU$(n > 2)$ case, fluctuations are responsible for the change of the order of the transition for SU$(2)$. Thus, further analysis of the effect of fluctuations is needed but it is beyond the  scope of the present work.   

 For $a_s < a^c_s$ the three energy has one minimum located at $\bar{m}_0 = 0$. However, for $a_s >  a_c$ 
the free  energy exhibits tree degenerate minima, corresponding (in `cartesian' $(m_3,m_8)$ coordinates)  to $\mathbbm{M}_0 \propto (0,-1)\: \bar{m}_0$ and $\mathbbm{M}_0 \propto  (\pm \sqrt{3}/2,  1/2)\: \bar{m}_0$. However, it needs to be noticed that these three minima represent the same physical state,  as the result of the invariance of the free energy under the transformation $\theta \to \theta + 2\pi j/3$, where $j = 1, 2$, which corresponds to a cyclic permutation of the SU(3) indices   $1 \rightarrow 2 \rightarrow 3 \rightarrow 1$, or in other words, to the existence of three
(non-commuting) SU$(2)$ subalgebras in SU$(3)$, into which the larger group $SU(3)$ can be spontaneously broken. Thus,
let us choose  $\mathbbm{M}_0  \propto - \bar{m}_0\:  \mathbbm{T}^{8} = - \bar{m}_0 \: ( \mathbbm{1} - 3\, \mathbf{e} \otimes \mathbf{e}^{\dag})$  (corresponding to $m_3 = 0$ and $m_8  = - m_0$), where  $\mathbf{e}^{\dag}= (0,0,1)$.   This ferromagnetic state corresponds to a gas where one of the species Fermi surface  ($\alpha = 3$, in this case) grows at the expense of the two others, which remain degenerate. 
This state left invariant under the transformations generated by the SU$(2)$ subalgebra span by $\{\mathbb{T}^1,  \mathbbm{T}^2, \mathbbm{T}^3 \}$. Furthermore, it is left invariant by the U$(1)$ transformations generated by $\mathbbm{T}^8$. Thus, the little group of transformations leaving the ground state invariant  is  $H = \mathrm{SU}(3)\otimes \mathrm{U}(1)$.  
If the interaction is increased further on, the remaining SU$(2)$ group may be also spontaneously broken down to U$(1)$ in a subsequent transition.

   Generally speaking, unlike SU$(2)$ case,  the SU$(n)$ may be spontaneously broken in a cascade of phase transitions. A general analysis of the possibilities can be given by considering the structure of the order parameter. As pointed out above, the order parameter is a traceless hermitian matrix,  $\mathbbm{M}= \sum_{a}\mathbbm{m}_{a} \mathbbm{T}^a$, which
transforms according to the adjoint representation of SU$(n)$. Thus, when diagonalized, it
has $n-1$ independent eigenvalues. If  only $k < n$ of them turn out to be equal, the  symmetry breaking
pattern (up to discrete groups) will be  $\mathrm{SU}(n) \to \mathrm{SU}(k)\times \left[\mathrm{U}(1)\right]^{n-k}$. Another more symmetric
state occurs when there are only two distinct eigenvalues and hence  $\mathrm{SU}(n) \to \mathrm{SU}(n-k)\times 
\mathrm{U}(k)$ ($k \leq n/2$). When all the $n-1$ eigenvalues turn out to be different, $\mathrm{SU}(n) \to \left[\mathrm{U}(1)\right]^{n-1}$, 
etc.  A simple example of the broken symmetry ground states (at the Hartree-Fock level) is provided by  the state 
$|\Phi(p^1_\mathrm{F}, \ldots, p^{n}_\mathrm{F})\rangle = 
\prod_{\alpha=1}^n\prod_{|\mathbf{p}| < p^{\alpha}_\mathrm{F}} c^{\dag}_{\alpha}(\mathbf{p}) |0\rangle,$
where $|0\rangle$ is the particle vacuum.  
The number of different eigenvalues tell us how many of the Fermi momenta coincide.
More generally, any FM ground state can be considered to be adiabatically connected
with an SU$(n)$ rotation of $|\Phi(p^1_\mathrm{F}, \ldots, p^{n}_\mathrm{F})\rangle$.
When there are only two different eigenvalues, the order parameter manifold 
$\mathcal{M} = G_{n,k}  = \mathrm{SU}(n)/[SU(n-k)\times \mathrm{U}(k)]$,
that is,  a Grassmanian manifold~\cite{qhferromagnet}. In particular, for 
$k = 1$, $G_{n,1} \simeq \mathbbm{CP}^{n-1}$, the complex projective space.  These
manifolds have non-trivial second homotopy group,  $\pi_2(G_{n,k}) = \mathbbm{Z}$
($n \geq 2$), which implies that  these  FM  phases 
can sustain topologically stable excitations that
are skyrmions  in $d = 2$ and monopoles in $d = 3$.
Furthermore, when SU$(n)$ breaks into a subgroup  containing more than one $U(1)$,
$\pi_2(\mathcal{M}) = \mathbbm{Z}^p$, where $p \leq n-1$ is the total number of $U(1)$'s. 
The corresponding phases thus support complex types of topological defects
described by several (integer) topological charges

 Finally, let us mention that as far as the experimental realization of SU$(n > 2)$ ferromagnetism is concerned,  the above discussion suggest that the most convenient approach to observe an SU$(n > 2)$ paramagnet to  ferromagnet phase transition in the $^{173}$Yb system  is to increase the scattering length by means of an optical Feshbach resonance~\cite{optical_feshbach}. Ferromagnetism
may also appear when the system is loaded in an optical lattice. However, this phase will compete with others (see next section)
and further analysis will be required to understand the full phase diagram of the lattice system. 

 \begin{center}
\begin{figure}[th]
\includegraphics[width= \columnwidth]{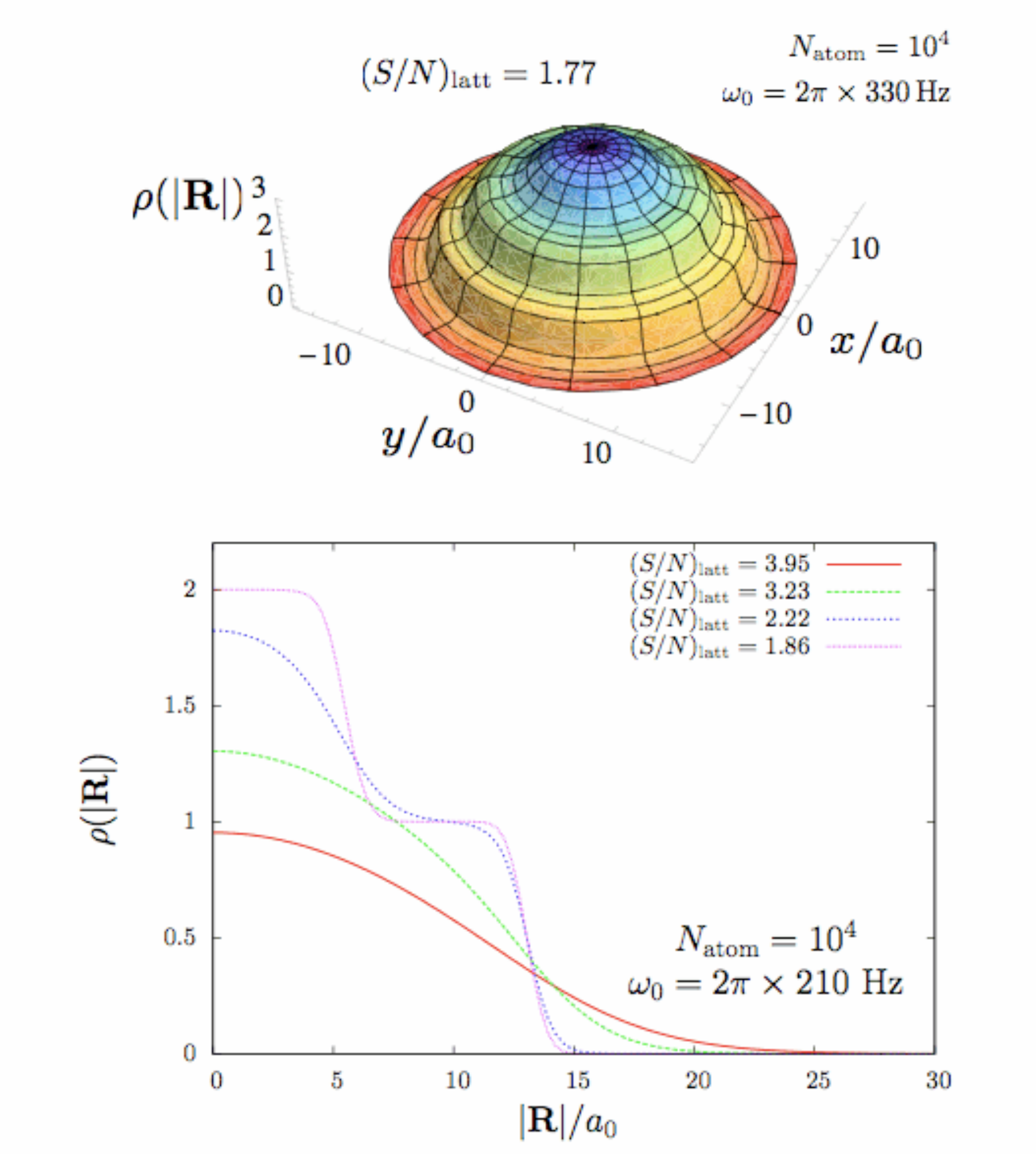}
\caption{Site occupation, $\rho(|{\bf R}|)$, as a function of the distance 
to the center of the cloud, $|{\bf R}|$, in units of the lattice parameter $a_0$,
obtained from an SU$(6)$ Hubbard model on a cubic lattice in $d=3$. 
We have used parameters similar to those of current experiments~\cite{Kyoto_exp,Takahashi_private}. 
In both the upper and lower panel $U/t \simeq 45$ (lattice depth $10\, E_R$). 
The upper panel shows the emergence of a
half-filled ($\rho(|{\bf R}|) = 3$) region near the center.
The lower panel shows the emergence of the
``Mott shell'' structure as the lattice entropy per atom is reduced. The red curve corresponds to 
the currently achievable temperatures of the gas (before \emph{adiabatically} ramping up the  lattice) of
$T_{\mathrm{init}}/T_F = 0.4$~\cite{Kyoto_exp,Takahashi_private} 
($T_F$ being the Fermi temperature of the harmonically trapped gas). The entropy is estimated using 
$(S/N)_{\mathrm{latt}}  = \pi^2 (T_{\mathrm{init}}/T_F)$, for a trapped
non-interacting gas.  For the smallest
value of the entropy, $(S/N)_{\mathrm{latt}} = 1.86$, the lattice temperature becomes comparable to
the hopping (and therefore the atomic approximation breaks down).    
Hopping will further reduce the occupation near the edges of the cloud.}
\label{fig:fig1}
\end{figure}
\end{center}
\section{$^{173}$Yb gases in an optical lattice: SU$(n=6)$ Hubbard model}\label{sec:HM}

  When the system is loaded in an optical lattice other phases may become more energetically favorable.
In a uniformly filled lattice, as the filling  $\nu  = \mathcal{N}/\mathcal{M}$ 
 ($\mathcal{N}$ and $\mathcal{M}$ being the total atom and site numbers, respectively) approaches
 half-filling,  $\nu = \frac{n}{2}$ other phases can become more favorable than ferromagnetism.
Let us assume that the $^{173}$Yb loaded in a  lattice are accurately described by 
 a single-band Hubbard model, $H = \sum_{\mathbf{p}} \epsilon_0(\mathbf{p}) \: c^{\dag}_{\alpha}(\mathbf{p})c^{\alpha}(\mathbf{p})  + \frac{U}{2} \sum_{\mathbf{R}} \left[\rho(\mathbf{R})\right]^2$,
 where $\epsilon_0(\mathbf{p}) = - 2 t \sum_{i=1}^d \cos k_i a_0$ is the free particle dispersion ($a_0$ is the lattice parameter), and $\rho(\mathbf{R}) = n^{\alpha}_{\alpha}(\mathbf{R})$ the total site occupancy.  This model will be accurate when the 
 lattice is sufficiently deep.  If the lattice depth is further increased so that hopping is suppressed along one direction, 
 the system dimensionality will become effectively  $d = 2$. In this case,  there is strong evidence
 that for large values of $n$~\cite{Marston,honerkamp_sun}, a staggered flux 
 phase~\cite{SF_papers} with atom currents circulating in opposite directions in neighboring 
 plaquettes  will be favored as the ground state. This phase breaks time-reversal
 as well as lattice translation symmetries but does not break SU$(n)$:
 thus the system will exhibit long-range order also at finite temperatures in $d = 2$.
 In Ref.~\cite{honerkamp_sun} it was argued that $n = 6$ is indeed a borderline case where this
 phase competes with a flavor density wave that breaks both lattice translation symmetry 
 and SU$(n)$ symmetry. The order parameter of this phase, 
 $D^{\alpha}_{\beta}({\bf Q}) = \frac{1}{\Omega} \sum_{\mathbf{p}} \langle 
 c^{\dag}_{\beta}(\mathbf{p}) c^{\alpha}(\mathbf{p}+{\bf Q})\rangle $ 
 (where ${\bf Q} = (\pi, \pi)$ at half-filling),
 is also a tensor belonging to the adjoint representaiton. Thus, if
 $\mathrm{Tr} \left[ \mathbbm{D}(\mathbf{
 Q}) \mathbbm{T}^{a}  \right] =  D^{\alpha}_{\beta}({\bf Q}) (\mathbbm{T}^a)_{\alpha}^{\beta} \neq 0$  
 for any $a = 1, \ldots, n$,  the SU$(n)$ symmetry will be broken in one of the
 the same patterns as in the FM case. 
 Thus, the $^{173}$Yb gas in an optical lattice may be an ideal system to study
 the rich phase diagram resulting from the competition of all these phases.

However, the temperatures that are currently achievable in an optical lattice (typically larger 
than the hopping amplitude $t$, see caption on Fig.~\ref{fig:fig1})  are  well above the temperature  scales where the 
ordered phases discussed above may  occur.  Furthermore, the presence of the harmonic trap leads to an  \emph{inhomogeneous} filling  of the 
lattice. The variation of the site occupation across the trap can be  estimated  in the 
so-called atomic (\emph{i.e.} $t = 0$) limit of the SU$(n)$ Hubbard model introduced 
above upon including the harmonic trap: $H_{\rm at} = \frac{U}{2}  \sum_{\mathbf{R}} \left[\rho(\mathbf{R})\right]^2 
+  V_\mathrm{t} \sum_{\mathbf{R}}   \left( \frac{\mathbf{R}}{a_0} \right)^2 \rho({\mathbf{R}})$,
where $V_{\mathbf{t}} = \frac{1}{2} m \omega^2_0 a^2_0$ is the trapping energy and
$a_0$ the (optical) lattice parameter.  The average site occupation can be thus obtained
from $\langle \rho(\mathbf{R})\rangle = 
\mathrm{Tr} \left[\rho_{\mathbf{R}} e^{-(H_{\mathrm{at}} - \mu N)/T} \right]/\mathrm{Tr} e^{-(H_{\mathrm{at}} - \mu N)/T} $,
 with $N = \sum_{\mathbf{R}} \rho({\bf R})$ and $T$ the absolute temperature. Hence,
\begin{equation}
\langle \rho(\mathbf{R}) \rangle  = T \frac{\partial}{\partial \mu} \ln \left[\displaystyle\sum_{p=0}^{n} 
\mathcal{C}^{n}_{p} e^{- \frac{U}{2T} p^2 - \frac{V_{\mathrm{t}}(\mathbf{R}/a_0)^2 - \mu}{T} p}\right],
\end{equation}
where $\mathcal{C}^{n}_{p} = \frac{n!}{p! (n-p)!}$ is  the energy degeneracy of a single-site state 
containing $p$ particles. The chemical potential $\mu$ must be adjusted to fix the total
number of particles and $T$ must be such that the entropy of the lattice equals
that of the gas before \emph{adiabatically} ramping up the lattice.  A plot of the 
site occupancy  as a function of the radial distance to the center of the trap $|{\bf R}|$
is displayed in Fig.~\ref{fig:fig1}.

\section{Summary and Conclusions}\label{sec:concl}

To sum up, by using  Fermi liquid theory, we have discussed  Fermi liquid (Pomeranchuk) instabilities
in the spin chanel of a strongly interacting ultracold $^{173}$Yb gas exhibiting an enlarged SU$(n = 6)$ symmetry.
Focusing on the Ferromagnetic instability, which does not break space rotation or translation symmetries, 
we have shown that the transition is generically first order (at least, at the mean level).  Such an instability
corresponds to a phase transition which can observed by increasing the scattering length using an optical
Feshbach resonance or/and in an optical lattice. For the continuum case, the first order of the
transition implies that the transition takes places at a slightly smaller  value of the scattering length
than the value provided by the Stoner criterion, which we find to be independent of the order of the group, $n$.  
Furthermore, using the smaller group SU$(3)$ as an example, 
we have illustrated how the larger unitary symmetry is broken by an explicitly analysis of the Landau
free energy derived from  the microscopic Hamiltonian. Thus, we found that SU$(3)$ is spontaneously broken down
to   SU$(2) \otimes $ U$(1)$. 

 On general symmetry grounds, we can expect a number of  symmetry-breaking patterns for SU$(6)$, 
which may be the result of  not just one but a cascade of phase transitions between ferromagnetic phases.
These SU$(n)$ ferromagnets system can sustain exotic topologically stable excitations, such as 
skyrmions in $d = 2$ and  monopoles in $d = 3$.  The resulting phase diagram may be indeed quite rich, and 
will be explored elsewhere~\cite{unpub}.    An interesting direction would be also to apply the analysis, based
on Hertz theory~\cite{Hertz06}, to study other Fermi surface instabilities in SU$(n)$ spin channel or to the  
flavor density wave of Ref.~\cite{honerkamp_sun} on the lattice. Based on the group theoretic properties 
of the order parameter, the latter may also turn out to be first order
at the mean field level.  Furthermore, in the optical lattice, the $^{173}$Yb system also offers other possibilites. 
such a realization of the  staggered flux phase, which breaks the lattice  but not SU$(n)$ symmetry. 
However, under current experimental conditions, the temperature of the gas in the lattice is well above the 
ordering temperature for these phases.  In this limit, we have obtained the density
profile in a harmonic trap (see Fig.~\ref{fig:fig1}).

  The fact that the spin of the Ytterbium atom in its ground state is entirely nuclear implies that
its coupling to a real magnetic field is very weak and this renders magnetic fields impractical 
to detect the population of different species.  However, the ferromagnetic phases and the topological 
defects discussed above  could be detected by means of the optical Stern-Gerlach 
effect induced by off-resonant circularly
polarized light~\cite{Takahashi_private}. Although this method has not been yet demonstrated
experimentally, it provides the most direct way to image the population of each species in 
a single shot measurement~\cite{Takahashi_private}.  
Nevertheless, we hope that the possibilities discussed above for the observation 
of new and exotic many-body states in the $^{173}$Yb will spur further theoretical and
experimental research along these lines. The first step in this direction may be measuring the
site occupation in an optical lattice, which can be carried out as explained in Ref.~\onlinecite{Ytterbium_mott}. 

 We are indebted to Y. Takahasi and D.~K.~K. Lee for fruitful discussions. MAC acknowledges 
financial support of the Spanish MEC through grant No. FIS2007-066711-C02-02 and
CSIC through grant No. PIE 200760/007. AFH acknowledges financial support from 
EPSRC(UK) through grant EP/D070082/1.

{\bf Note added:} After completion of our work, we became aware of the work by Gorshkov 
et al.~\cite{Gorshkov}, who also pointed out enlarged SU(n) symmetries of alkaline-earth atomic gases.

 \end{document}